\documentclass[
	aps,
	prl,
	preprint,
	superscriptaddress,
	showpacs
]{revtex4-1}
\usepackage{graphicx}
\usepackage{amssymb,amsmath,amsfonts, latexsym}
\usepackage[sort&compress]{natbib}
\usepackage{mathrsfs} 
\usepackage[dvipsnames]{xcolor}
\usepackage{multirow}
\usepackage{longtable}
\usepackage{hyperref}\hypersetup{colorlinks=true,linkcolor=blue,citecolor=blue}
\usepackage{bm}
\newcommand{\tens}[1]{\mathbf{\boldsymbol{#1}}}
\renewcommand{\vec}[1]{\mathbf{\boldsymbol{#1}}}
\usepackage{times, mathptmx}
\usepackage{CJKutf8}

\definecolor{light-gray}{gray}{0.5}
\begin{document}


\begin{CJK*}{UTF8}{gbsn}
\title{Determination of the stretch tensor for structural transformations}
\author{Xian Chen (陈弦)}
\email[]{xianchen@lbl.gov}
\affiliation{Aerospace Engineering and Mechanics, University of Minnesota, Minneapolis, MN 55455 USA}
\affiliation{Advanced Light Source, Lawrence Berkeley National Lab, CA 94702 USA}
\author{Yintao Song (宋寅韬)}
\affiliation{Aerospace Engineering and Mechanics, University of Minnesota, Minneapolis, MN 55455 USA}
\author{Nobumichi Tamura}
\affiliation{Advanced Light Source, Lawrence Berkeley National Lab, CA 94702 USA}
\author{Richard D. James}
\affiliation{Aerospace Engineering and Mechanics, University of Minnesota, Minneapolis, MN 55455 USA}


\date{\today}

\begin{abstract}
The {\it transformation stretch tensor} plays an essential role in the evaluation
of conditions of compatibility between phases and the use of the Cauchy-Born rule.
This tensor is difficult to measure directly from experiment.
We give an algorithm for the determination of the transformation stretch tensor
from x-ray measurements of structure and lattice parameters.  When evaluated
on some traditional and emerging phase transformations the algorithm gives
unexpected results.
\end{abstract}

\pacs{\textcolor{red}{61.50.Ks}}

\maketitle
\end{CJK*}
The structural transformations commonly occur in application of functional materials. Typical examples of phase transformation driven phenomena include shape memory alloys, ferroelectricity, piezoelectricity, colossal magnetoresistance and superconductivity. It has been demonstrated that material reliability depends, essentially, on the reversibility of the transformation. It is therefore important to understand how reversibility can be achieved and how the transformation occurs at the lattice and atomic level.
The \emph{transformation stretch tensor}, $\mathbf U$, 
is the stretch part of the linear transformation that maps the crystal structure
from its initial phase to the final phase \cite{Ball_1987, Ball_1992, Bhattacharya2003, Chen_2011}.  
Recently, the reversibility, the thermal hysteresis, and the 
resistance to cyclic degradation of functional materials have been linked 
to properties of the transformation stretch tensor.   For example, when the middle eigenvalue $\lambda_2$
of $\mathbf U$ is tuned to the value 1 by compositional changes, the measured width of the thermal hysteresis loop drops
precipitously to near 0 in diverse alloy systems \cite{Cui_2006, Zarnetta_2010, Chen_2013}.  Assuming the Cauchy-Born rule for martensitic materials \cite{Bhattacharya2003, Ericksen_2008, Pitteri_2010}, the condition $\lambda_2 = 1$ implies a special condition of compatibility between phases by
which the undistorted austenite phase and a single undistorted variant of the martensite phase fit perfectly together
at an interface.  Even stronger conditions of compatibility known as the {\it cofactor conditions}
($\lambda_2 = 1$ together with either $|\mathbf U^{-1} \mathbf e| = 1$ or $|\mathbf U \mathbf e| = 1$, where
$\mathbf e$ is unit vector on a 2-fold symmetry axis of austenite), lead to even lower hysteresis and significantly 
enhanced reversibility during cyclic transformation \cite{Song_2013}. $\mathbf U$ also plays an
important role in determining the elastically favored orientations of precipitates for  
diffusional transformations \cite{Voorhees_1988, Chen_2011}. 
	
\begin{figure}[htp]
\centering
\includegraphics[width=5in]{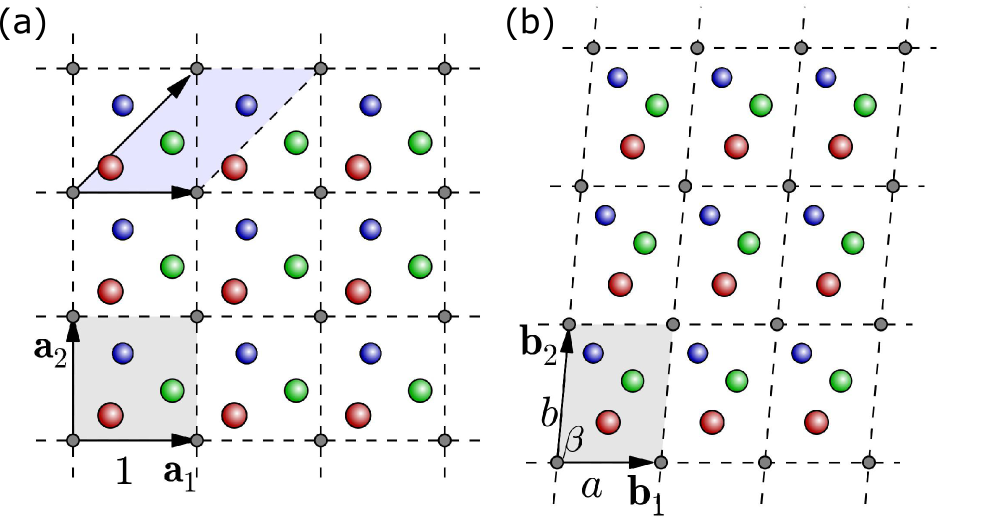}
\caption{Non-uniqueness of Cauchy-Born deformation gradient from (a) square lattice to (b) oblique lattice. Red, blue and green balls represent different atomic species. Gray dots are lattice points}\label{CB}
\end{figure}
	
In principle, the determination of the stretch tensor $\mathbf U$ for a structural transformation
is straightforward. Suppose the primitive lattice vectors of initial and final phases are, respectively, 
linearly independent vectors 
$\mathbf a_i$ and $\mathbf b_i$ for $i = 1, 2, ...d$ where $d$ is the dimension of the lattice.
A nonsingular linear transformation $\mathbf F$ can be defined uniquely by
\begin{equation}\label{map}
\mathbf F \mathbf a_i = \mathbf b_i, \quad i = 1, 2, ...d,
\end{equation}
and the polar decomposition of $\mathbf F$ is written $\mathbf F = \mathbf Q \mathbf U$, where $\mathbf Q$
is orthogonal and $\mathbf U$ is positive-definite and 
symmetric, called the transformation stretch tensor.
The notation $\mathbf a_i  \to \mathbf b_i$ denotes the {\it lattice correspondence}. In the case of transformation in Fig. \ref{CB}, one choice of the lattice correspondence can be $\mathbf a_1 \to \mathbf b_1, \ \mathbf a_2 \to \mathbf b_2$ where $\mathbf a_1 = [1, 0]$, $\mathbf a_2 = [0, 1]$ and $\mathbf b_1 = [a, 0]$ and $\mathbf b_2 = [b \cos \beta, b \sin \beta]$.

As is well-known \cite{Wayman_1964, Bowles_1972}, $\mathbf F$ and $\mathbf U$ are not uniquely determined by the two lattices.  This
follows from the fact that there are infinitely many choices of lattice correspondence.
From  Fig.~\ref{CB}, the alternative set of vectors $\mathbf a_1$ and $\mathbf a_1 + \mathbf a_2$ describes the same  lattice (a), which results in a different correspondence from (a) to (b). This obviously changes the $\mathbf F$ and thus the transformation stretch tensor $\mathbf U$. More generally, any two sets of primitive lattice vectors for a given lattice are related by a 
{\it lattice invariant transformation} \cite{Ball_1992} i.e., a unimodular matrix of integers. If we allow an invariant transformation for both initial and final phases,
the ambiguity of $\mathbf F$ is $\mathbf F \to \Lambda_\text{(f)} \mathbf F \Lambda_\text{(i)}^{-1}$ where $\Lambda_\text{(i)}$ and $\Lambda_\text{(f)} $ denote the lattice invariant transformation for initial and final lattices, respectively. 

The linear transformation $\mathbf F$ represents the change of periodicity of the two phases.
The individual atoms denoted by the red, blue and green balls in Fig.~\ref{CB} may shuffle in
various ways, giving rise to different space group symmetries, but it is the linear transformation
$\mathbf F$ that relates to macroscopic deformation and therefore to conditions of
compatibility \citep{Bowles_1954, lieberman_57, Wayman_1964, Knowles_1981, Ericksen_1984, Ball_1987, Ericksen_2008, Pitteri_2010, Zhang_2009, Chen_2013}.  This idea is formalized by the
weak Cauchy-Born rule \cite{Ericksen_2008, Bhattacharya_2004}.   This rule is used to define the dependence
on deformation of the free energy at
continuum scale from the free energy density at atomistic scale for complex lattices with multiple atoms per unit cell and inhomogeneous deformations. Inhomogeneous deformations $\mathbf y (\mathbf x)$ locally satisfy the same rule as above:
$\mathbf b_i = \nabla \mathbf y\, \mathbf a_i$, where $\mathbf a_i$ and $\mathbf b_i$ represent 
the local periodicity.  Note that we use a geometrically exact description
here.  A geometrically linear description (i.e., as in linear elasticity) would not be sufficiently accurate to
describe transformations here for the purposes of imposing the conditions of compatibility
(see \cite{Chen_2013} for calculations of the error in various cases).

\begin{figure}[htp]
\centering
\includegraphics[width=4in]{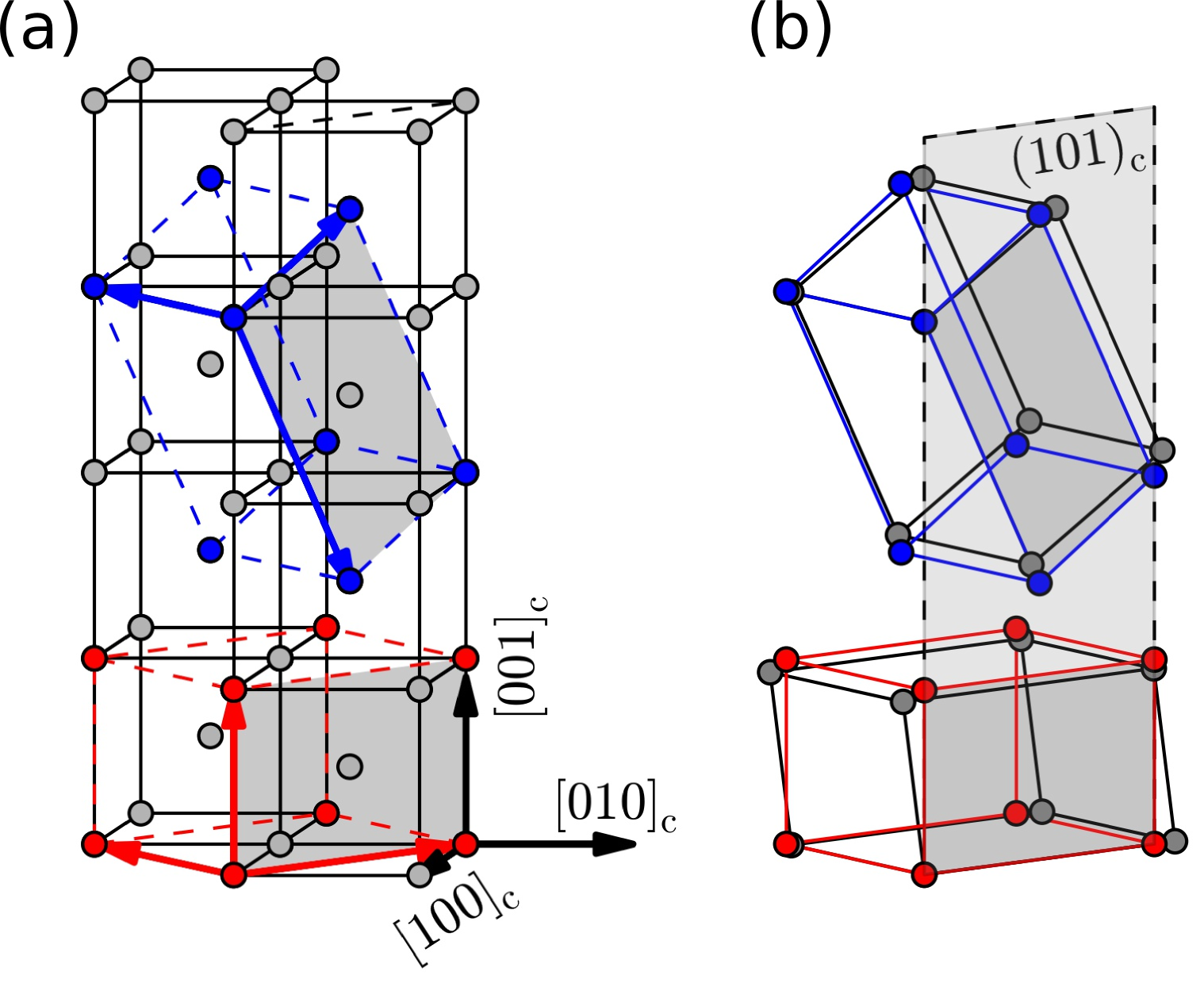}
\caption{\label{notBain}
The least atomic movements during the structural transformation. 
(a) The bcc lattice and two of its sublattices (red and blue) of size 4.
(b) Comparison between these bcc sublattice unit cells 
and the primitive cell of the final phase (gray; for clarity
atoms in the unit cell are not shown). 
}
\end{figure}

Based on a natural intuition that ``a mode of atomic shift requires minimum motion" \citep{Bain_1924}, Bain proposed a famous lattice correspondence in 1924 for the formation of bcc $\alpha$Fe from fcc $\gamma$Fe . The correspondence has been well-accepted and applied to study numerous phase transformations \cite{Tadaki_1970, Otsuka_1971, Knowles_1980, Chu_1993,
Ball_1992, Bhattacharya2003, Huang_2003}. 
To illustrate how easy the Bain correspondence misses the smallest strain, we construct an example of transformation from a bcc lattice with $a_0 = 1$ to a monoclinic lattice with $a = 0.961$, $b = 1.363$, $c = 1.541$, and $\beta = 97.78^\circ$. 
Fig.~\ref{notBain}(a) shows the bcc lattice with two sublattice unit cells (red and blue). 
Conventional wisdom would say that the Bain
correspondence (red $\to$ gray in Fig.~\ref{notBain}(b), bottom) 
is appropriate for this transformation.  However, our algorithm proposed later in this letter reveals an
unexpected alternative correspondence (blue $\to$ gray,  Fig.~\ref{notBain} (b), top).  
Both contain 4 lattice points ($n=4$) in the unit cells, and the shape 
and size of them are similar to the primitive cell of monoclinic lattice. 
Fig.~\ref{notBain}(b) shows the comparison of distortions for both transformation mechanisms.
Notice that both mechanisms give exactly the same final monoclinic lattice.
However, by quantitative calculation,
the principle strains for the new correspondence are in fact smaller than those for the Bain correspondence.

The significance of finding the correct lattice correspondence for structural phase transformations is 
emphasized in the literature \cite{Wayman_1964, Bowles_1972}. 
The problem was well-appreciated by Lomer \cite{Lomer_1955} as early 
as the mid-1950s.  In his study of the mechanism of
the $\beta \to \alpha$ phase transformation of U$_{98.6}$Cr$_{1.4}$, he examined theoretically 
(by hand) 1,600 possible transformation mechanisms, and reduced this to three correspondences
having the smallest principle strains, which he considered the likely candidates. 


Direct experimental measurement of the macroscopic finite strain of transformation, 
together with accurate structural characterization by X-ray diffraction provides a possible way to determine the lattice correspondence and thus the transformation stretch tensor. But this is technically difficult due to (i) the need
for an oriented single crystal, (ii) the need to remove the inevitable fine microstructures
that form during transformation due to constraints of compatibility, and (iii) the need for an accurate measure of full finite strain tensor along known crystallographic directions.    
We also noticed that using a state-of-art high resolution TEM on a 
pre-oriented single crystal sample can not definitively remove the ambiguities among many
lattice correspondences due to some inevitable obstacles: tracking the evolution of 
diffraction spots in a fast structural transformation process, 
simultaneously indexing both phases, and most significantly, 
finding a special zone that can unambiguously reveal the differences among various 
lattice correspondences. 

In this letter we propose an algorithmic approach to search the $N$ best choices of
lattice correspondence for a structural transformation, by minimizing a particular 
strain measure between initial and final lattices.  
The input to the algorithm is the underlying periodicities (the remaining space group information is not needed) 
and the lattice constants of the two phases. 
The output from the algorithm is the $N$ best choices of lattice correspondence and the associated transformation stretch tensors. Users can customize how many solutions they like by manipulating $N$.  The results can be used as a reference by the advanced structural characterization facilities for the determination of orientation relationships,
and it can be integrated with first principles calculations to give starting
points for the determination of energy barriers or interfacial distortion profiles.

Consider a Bravais lattice ${\cal L} = \{ \sum n^i \mathbf e_i : n^1, \dots n^d \in \mathbb Z^d \}$ determined by
linearly independent {\it lattice vectors} $\mathbf e_1, \dots, \mathbf e_d \in\mathbb R^d$, $i=1,\ldots,d$, and assemble
the lattice vectors as the columns of a $d\times d$  matrix $\mathbf E = (\mathbf e_1, \ldots, \mathbf e_d)$. 
$\cal L$ can equivalently be denoted 
\[
{\cal L} = {\cal L}(\mathbf E) = \big\{ \mathbf r\in\mathbb R^d: \mathbf r = \mathbf E\vec\xi, \vec\xi \in \mathbb Z^d \big\}.
\]
Without loss of generality,  by switching the sign of $\mathbf e_1$ if necessary, we assume that 
$\det \mathbf E > 0$. This determinant is the ($d$-dimensional) volume of a unit cell of 
${\cal L}(\mathbf E)$. 

Given two lattices ${\cal L}(\mathbf E)$ and ${\cal L}(\mathbf E')$, the $d\times d$ nonsingular
matrix $\mathbf L$ satisfying $\mathbf E' = \mathbf E \mathbf L$ is called 
the \emph{correspondence matrix} from ${\cal L}(\mathbf E)$ to ${\cal L}(\mathbf E')$.
As noted above, the two lattices ${\cal L}(\mathbf E)$ and ${\cal L}(\mathbf E')$ are the same if and only if 
the correspondence matrix $\mathbf L$ is a unimodular matrix of integers, or, briefly, 
$\mathbf L \in GL(d, \mathbb Z)$.
If a correspondence matrix $\mathbf L$ is a matrix of integers with $|\det \mathbf L| > 1$, then
${\cal L}(\mathbf E')$ is a \emph{sublattice} of ${\cal L}(\mathbf E)$.  The quantity
$|\det \mathbf L|$ is the volume ratio of the unit cell of $\mathcal L(\mathbf E')$
to that of $\mathcal L(\mathbf E)$.

Correspondence matrices are often reported for conventional rather than
primitive descriptions, particularly for  7 of the 14 types of Bravais lattices in 3D.  
For example, the conventional description for an fcc lattice with 
lattice parameter $a_0$ is an orthogonal basis, so
 $\mathbf E_{\rm conv} = a_0\mathbf I = \mathbf E \tens\chi$,
where, for example,
\[
\mathbf E = \frac{a_0}{2}\begin{bmatrix}1&0&1\\1&1&0\\0&1&1\end{bmatrix},~~
\tens\chi = \begin{bmatrix}1&1&-1\\-1&1&1\\1&-1&1\end{bmatrix}.
\]
Here, $\det \tens \chi = 4$ so the volume of the conventional unit cell
is 4 times that of the primitive cell. 
From now on, the symbol $\tens\chi$ is reserved for a correspondence matrix 
from the primitive to conventional unit cell of a Bravais lattice:
$\mathbf E_{\rm conv} = \mathbf E \tens\chi$. 

We  seek a sublattice of $\mathcal L (\mathbf E_A)$ that is mapped to the
primitive lattice of $\mathcal L (\mathbf E_B)$.   (The algorithm
can easily handle the case in which we take sublattices of both lattices.)
As above, let $\mathbf E_A = (\mathbf a_1, ... , \mathbf a_d)$ and 
$\mathbf E_B = (\mathbf b_1, ..., \mathbf b_d)$.
Let $\tens\ell \in \mathbb Z^{d\times d}$, $\det \tens\ell > 0$, be the correspondence matrix giving the sublattice $\mathcal L(\mathbf E_A \tens \ell)$ that is mapped to the final lattice $\mathcal L(\mathbf E_B)$ during the transformation. 
The basic equation \eqref{map} in this case becomes
$\mathbf F \mathbf E_A \tens\ell= \mathbf E_B $, and the transformation stretch tensor
$\mathbf U$ is the unique positive-definite square root of $\mathbf F^T \mathbf F$. 

We introduce the following function as a measure of the
distance from $\mathbf U$ to $\mathbf I$:
\begin{equation} \label{distance}
\begin{split}
\text{dist} (\tens\ell, \tens E_A, \tens E_B) & = \left\Vert (\mathbf F^T\mathbf F)^{-1} - \mathbf I \right\Vert^2 \\
& = \left\Vert \tens E_A\tens \ell\tens E_B^{-1}\tens E_B^{-T} \tens\ell^T \tens E_A^T - \tens I\right\Vert^2.
\end{split}
\end{equation}
$\Vert \cdot \Vert$ denotes the Euclidean norm, $\Vert \tens A \Vert = \sqrt{\text{tr\,}\tens A^T \tens A}$. 
The distance  \eqref{distance} is independent of rigid rotations of both lattices, and is particularly attractive from the point of view of symmetry. Physically, it represents the Lagrangian strain of the structural transformation. The use of inverse of $\mathbf F^T \mathbf F$ avoids possible noninvertibility of $\tens\ell$ that may arise during the minimization process. In addition, this norm is exactly preserved by point group transformations of both Bravais lattices.
That is,  if orthogonal tensors
$\tens R_A$ and $\tens R_B$ are, respectively, in the point groups of $\mathcal L(\mathbf E_A)$ and
$\mathcal L(\mathbf E_B)$, i.e., $\mathcal L(\mathbf E_A) = \mathcal L(\mathbf R_A \mathbf E_A)$ and
$\mathcal L(\mathbf E_B) = \mathcal L(\mathbf R_B \mathbf E_B)$, which, by the above implies that 
there exist associated matrices $\tens \mu_A$ and $\tens \mu_B$  such that 
$\mathbf R_A \mathbf E_A =\mathbf E_A \tens \mu_A$ and
$\mathbf R_B \mathbf E_B =\mathbf E_B \tens \mu_B$  then the distance transforms as
\begin{equation}\label{degenerate}
\text{dist} (\tens\mu_A \tens \ell \tens\mu_B,  \tens E_A,  \tens E_B) = \text{dist} (\tens\ell, \tens E_A, \tens E_B).
\end{equation}
Note that $\tens \mu_{A,B}$ are integral matrices of determinant $\pm 1$, so $\det \tens\ell = \det \tens\mu_A \tens \ell \tens\mu_B$.  Thus, immediately one minimizer of the distance with assigned determinant gives the expected
symmetry-related minimizers.  Physically, in the typical case of a symmetry-lowering 
transformation, e.g. the martensitic transformation, the distance function  \eqref{distance} automatically gives the equi-minimizing
variants of martensite.


As noted above it is typical to report the correspondence matrix in terms of the conventional basis 
instead of the primitive one. If $\tens\ell^*$ is a minimizer of 
$\text{dist} (\tens\ell, \tens E_A, \tens E_B)$ the conversion is done by
$\mathbf L^* = \tens\chi_A^{-1} \tens\ell ^* \tens\chi_B$. 
Note that $\mathbf L^*$ is not necessarily a matrix of integers.

A significant property of the distance function \eqref{distance} will be used to justify our algorithm below. Fixing $\mathbf E_A$ and $\mathbf E_B$, the distance function can be trivially extended to a function over
real matrices, $f(\mathbf L) = {\rm dist}(\mathbf L, \mathbf E_A, \mathbf E_B)$. Denoting
$\mathbf X_L = \mathbf E_A \mathbf L\mathbf E_B^{-1}\tens E_B^{-T} \mathbf L^T \tens E_A^T$
and using $\mathbf X_L \cdot \mathbf I \le \left\Vert \mathbf X_L \right\Vert \left\Vert \mathbf I \right\Vert
= \sqrt{3} \left\Vert \mathbf X_L \right\Vert$, we have
\begin{equation}
\begin{split}\label{ub}
f(\mathbf L) & = \left\Vert  \mathbf X_L \right\Vert ^ 2 - 2 \mathbf X_L  \cdot \tens I + 3\\
& \geqslant  \left\Vert  \mathbf X_L  \right\Vert ^ 2 -2 \sqrt{3}\left\Vert  \mathbf X_L  \right\Vert + 3
= (\left\Vert  \mathbf X_L  \right\Vert - \sqrt{3})^2,
\end{split} 
\end{equation}
Choose any integral matrix $\tens\ell_1$ and define $C_1 = f(\tens\ell_1)$.  By (\ref{ub})
the minimizer(s) of $f(\mathbf L)$ necessarily lie in the bounded set $\left\Vert  \mathbf X_L \right\Vert \le \sqrt{3} + \sqrt{C_1}$, that is,
$
 \left\Vert  \mathbf X_L \right\Vert ^ 2 \le 3 + C_1 +2 \sqrt{3 C_1}.
$
Let $\alpha$ be the minimum of $\Vert \mathbf X_L \Vert^2$ under the constraint $\Vert \mathbf L \Vert = 1$, 
then we have
\begin{equation}\label{radius}
\alpha\Vert \mathbf L \Vert ^ 4 \leqslant \left\Vert \mathbf X_L \right\Vert ^ 2 < 3 + C_1 +2 \sqrt{3 C_1}.
\end{equation}
That is, all the $\mathbf L$'s such that $f(\mathbf L) < C$ live in the sphere with the radius of 
$((3 + C_1 +2 \sqrt{3 C_1}) / \alpha)^{1/4}$ in $\mathbb R^9$.

Here is a brief outline of the algorithm for the determination of the $N$ best transformation stretch tensors and their associated lattice correspondences:
\begin{enumerate}

\item Calculate the primitive bases and the transformation matrices for the conventional cells from the input lattice parameters: $\mathbf E_\text{A, B}$ and $\tens\chi_\text{A, B}$. Calculate $\alpha$ by minimizing the term $\mathbf X_L$ with respect to $\mathbf L$ for all $\Vert \mathbf L \Vert = 1$.

\item Choose $N$ integral matrices ${\tens\ell_i}$, $i = 1,\ldots, N$ as the initial guess of the \emph{solution list} such that $\det \, \tens\ell_i$ is close to $\det\mathbf E_B / \det\mathbf E_A$ and ${\rm dist}(\tens\ell_i, \mathbf E_A, \mathbf E_B)$ is small.

\item \label{iteration} Let $C_1$ be the maximum $f(\tens\ell_i)$ for $\tens\ell_i$'s in the solution list. 

\item Calculate the distance for all integral matrices in the sphere of radius of $((3 + C_1 +2 \sqrt{3 C_1}) / \alpha)^{1/4}$. Update the solution list as necessary. If the solution list is changed, repeat from step \ref{iteration}.

\item For each solution $\tens\ell_i$, calculate the Cauchy-Born deformation gradient $\mathbf F_i = \mathbf E_B(\mathbf E_A \tens\ell_i)^{-1}$ and the transformation stretch tensor $\mathbf U_i = (\mathbf F_i^T\mathbf F_i)^{1/2}$. Finally, rewrite all the solutions in the conventional bases: $\mathbf L^*_i = \tens\chi_\text{A}^{-1}  \tens\ell_i \tens\chi_\text{B}$.

\end{enumerate}
	
\begin{figure}
\includegraphics[width=5.5in]{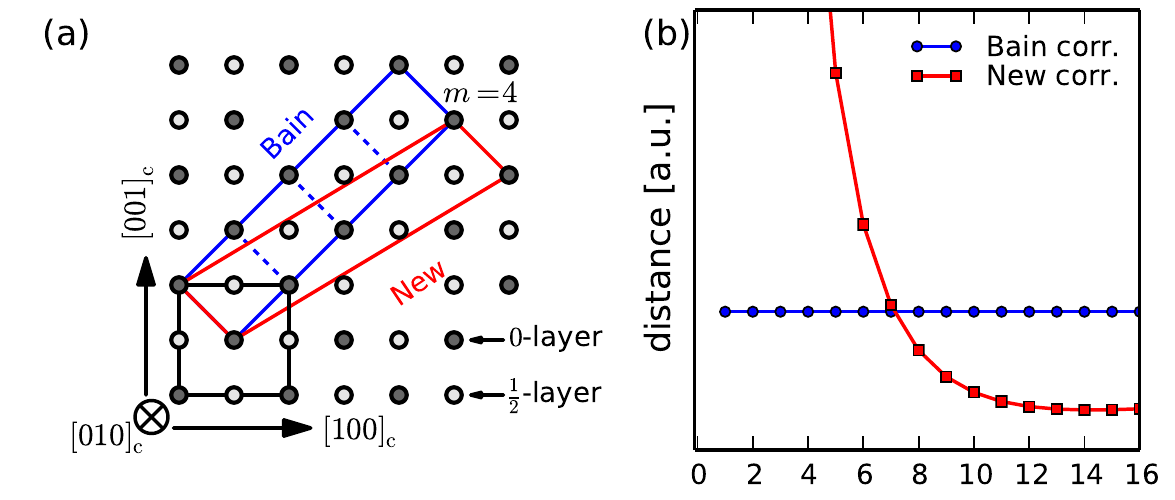}
\caption{Two possible lattice correspondences in an FCC to monoclinic transformation. (a) (010) projection of the FCC lattice: the dark (resp. light) atoms are in the $y=0$  (resp. $y = 1/2$) planes. The solid blue and red lines represent the two lattice correspondences respectively for $m=4$, where the the Bain correspondence
is in blue. The dashed blue lines indicate the modulation numbers $m=1,2, 3,4$. (b) shows the dependence of the values of the distance function on the modulation of the monoclinic $c$-axis for the two lattice correspondences. \label{modulation}}
\end{figure}

Note that the algorithm converges in a finite number of steps and gets all matrices with the $N$ lowest distances (up to the degeneracy in \eqref{degenerate}) because it searches through all matrices of integers satisfying the rigorous bounds (\ref{radius}).

In Fig.~\ref{modulation} we give an example computed by the algorithm that reveals a switch from Bain correspondence to a new correspondence with increasing lattice complexity.  Consider a transformation from an fcc lattice with lattice parameter $a_0=2$ to a monoclinic lattice with lattice parameters $a = 1.41, \ b = 1.99, \ c = 1.42\, m, \ \beta = 86^\circ$, where the integer $m>0$ denotes the modulation along monoclinic $c$-axis. Fig.~\ref{modulation}(a) shows the undeformed fcc lattice projected onto $(010)$ plane. The two correspondences given by the algorithm
are depicted for the $m=4$ case. 
Fig.~\ref{modulation}(b) shows the change in distance function for the two correspondences with $m$ varying from 1 to 16.  Initially Bain correspondence is much smaller than the new one, however it loses its privilege after the $7$th modulation. The results suggest that both kinds of lattice correspondence can be feasible in a structural transformation for some special lattice parameters, and in this case $m=7$ has this special status. As mentioned above, these long stacking period structures are common in martensitic phase transformations.

Table \ref{tab1} shows the results calculated by the algorithm for six materials. The types of transformation are diverse and the principle stretches are consistent with the references.  Among these examples, we list two solutions for Zn$_{45}$Au$_{30}$Cu$_{25}$. The material has been recently found to satisfy the cofactor conditions (the 2 constraints on $\mathbf U$ explained in paragraph 1) \cite{Chen_2013}, which have been shown \cite{Song_2013} to promote unusually low thermal hysteresis ($\approx2^\circ$C) and enhanced reversibility, owing to a fluid-like flexible martensite microstructure.  It was believed  \cite{Song_2013} to transform by the second solution, Table \ref{tab1}. However, the first solution is the one having the smallest transformation strain.
Coincidentally, the new transformation stretch tensor also satisfies closely the cofactor conditions. 
To investigate this further, the same sample of Zn$_{45}$Au$_{30}$Cu$_{25}$ used in \cite{Song_2013} was 
characterized by synchrotron X-ray Laue microdiffraction. The experiment has been conducted on beamline 12.3.2 of the Advanced Light Source, Lawrence Berkeley National Laboratory. Details on the experimental setup can be found in \cite{Kunz_2009}.  The Laue patterns were collected continuously as heating/cooling through the transformation temperature. These patterns were analyzed and indexed using the XMAS software \cite{Tamura_2014}. The orientation relationships are determined as the closest parallelisms of the crystallographic planes and zone axes between the indexed Laue patterns of austenite and martensite respectively. They are $(206)_\text{a} ||(20\, \bar{34})_\text{m}$, $(204)_\text{a} || (10\, \bar{26})_\text{m}$, $[21\bar{1}]_\text{a} || [26\, \bar{9}\,1]_\text{m}$, $[010]_\text{a}||[010]_\text{m}$ and $[1\bar{1}0]_\text{a} || [8\bar{9}1]_\text{m}$ (see supplementary for indexed diffraction patterns).  However, this determination with accepted error bars does not definitively distinguish these two mechanisms, since these relationships are so close that one could imagine that both
mechanisms occur simultaneously in the material.

In addition to the reversible martensitic transformation, the algorithm is applicable to a wide range of phase transformations even if the initial and final crystal structures do not have a group/sub-group relation.  Examples are Ti$_{95}$Mn$_{5}$ and Sb$_2$Te$_3$/PbTe (Table \ref{tab1}). The algorithm can be also applied to organic materials 
when the molecular chains have sufficient periodicity. One extreme example is the polymorphic transformation 
between two triclinic lattices in terephthalic acid (see Table \ref{tab1}). In this case the calculated principle stretches agree well with the measured macroscopic deformation of the polymorphic transformation of this material. 

 \begin{table}
 \caption{Transformation principle stretches (p. s.), the associated lattice correspondences (lat. cor.) and derived orientation relationships (o. r.) for various phase-transforming materials\label{tab1}}
{\footnotesize
\begin{ruledtabular}
\centering
\begin{tabular}{c|c|c|c}
materials & p. s. &lat. cor.& derived o. r.\\
\hline
\multirow{6}{*}{
\begin{tabular}{c}
Zn$_{45}$Au$_{30}$Cu$_{25}$ \cite{Song_2013} \\
${\rm L}2_1 \to {\rm M18R}$ 
\end{tabular}
} 
& $0.9363$ & $[\frac{1}{2}0\frac{1}{2}]_{\text{L2}_1}\to[100]_\text{M}$ & $(204)_{\text{L2}_1}||(\bar{1}\,0\, 26)_\text{M}$\\
& $1.0017$ & $[010]_{\text{L2}_1}\to[010]_\text{M}$ & $[1\bar{1}0]_{\text{L2}_1}||[8\bar{9}1]_\text{M}$\\
& $1.0589$ & $[\bar{4}05]_{\text{L2}_1}\to[001]_\text{M}$ &$[21\bar{1}]_{\text{L2}_1}||[26\, \bar{9}\, 1]_\text{M}$ \\ \cline{2-4}
& $0.9363$ &$[\frac{\bar1}{2}0\frac{\bar1}{2}]_{\text{L2}_1}\to[100]_\text{M}$ &$(204)_{\text{L2}_1}||(\bar{1}\,0\, 27)_\text{M}$\\
& $1.0006$ &$[010]_{\text{L2}_1}\to[010]_\text{M}$&$[1\bar{1}0]_{\text{L2}_1}||[9\bar{9}1]_\text{M}$\\
& $1.0600$ &$[\frac{9}{2}0\frac{\bar9}{2}]_{\text{L2}_1}\to[001]_\text{M}$&$[21\bar{1}]_{\text{L2}_1}||[27\,\bar{9}\, 1]_\text{M}$\\
\hline
\multirow{3}{*}{
\begin{tabular}{c}
CuAl$_{30}$Ni$_4$ \cite{Otsuka_1974} \\
$\beta_1 \to \gamma'$
\end{tabular}
} 
& $0.9178$&$[\frac{1}{2}0\frac{1}{2}]_\text{A}\to[100]_\text{B}$&$(110)_{\beta_1}||(121)_{\gamma'}$ \\
& $1.0231$ & $[010]_\text{A}\to[010]_\text{B}$&$[1\bar{1}\bar{1}]_{\beta_1}||[2\bar{1}0]_{\gamma'}$\\
& $1.0619$ &$[\bar{\frac{1}{2}}0\frac{1}{2}]_\text{A}\to[001]_\text{B}$&\\
\hline
\multirow{3}{*}{
\begin{tabular}{c}
Ti$_{95}$Mn$_5$ \cite{Knowles_1981} \\
bcc $\to$ hexagonal
\end{tabular}
} 
& $0.9052$ &$[010]_{\bf c}\to[100]_{\bf h}$ &$(1\bar1\bar1)_{\bf c}||(214)_{\bf h}$\\
& $1.0164$ &$[\frac{\bar1}{2}\frac{1}{2}\frac{1}{2}]_{\bf c}\to[010]_{\bf h}$&$[\bar1\bar21]_{\bf c}||[20\bar1]_{\bf h}$\\
& $1.1086$ &$[101]_{\bf c}\to[001]_{\bf h}$&\\
\hline
\multirow{3}{*}{
\begin{tabular}{c}
Ru$_{50}$Nb$_{50}$ \cite{Fonda_1999} \\
$\beta' \to \beta'$
\end{tabular}
} 
& $0.9791$ &$[11\bar{2}]_{\beta'}\to[100]_{\beta''}$ &$(100)_{\beta'}||(111)_{\beta''}$\\
& $1.0024$ &$[1\bar10]_{\beta'}\to[010]_{\beta''}$&$[011]_{\beta'}||[1\bar10]_{\beta''}$\\
& $1.0169$ &$[11\bar1]_{\beta'}\to[001]_{\beta''}$&\\
\hline
\multirow{3}{*}{
\begin{tabular}{c}
Sb$_2$Te$_3$ / PbTe \cite{Chen_2011} \\
fcc $\to$ hexagonal
\end{tabular}
} 
& $0.9384$ &$[\frac{\bar{1}}{2}\frac{1}{2}0]_{\bf c}\to[100]_{\bf h}$ &$(\bar110)_{\bf c}||(010)_{\bf h}$\\
& $0.9384$ &$[0\frac{1}{2}\frac{1}{2}]_{\bf c}\to[010]_{\bf h}$&$[001]_{\bf c}||[\bar48\bar1]_{\bf h}$\\
& $1.0779$ &$[22\bar2]_{\bf c}\to[001]_{\bf h}$&\\
\hline
\multirow{3}{*}{
\begin{tabular}{c}
Terephthalic acid \cite{Bailey_1967} \\
triclinic I $\to$ triclinic II
\end{tabular}
} 
& $0.8244$ &$[0\bar1\bar2]_{\bf I}\to[100]_{\bf II}$ &$[100]_{\bf I}||[112]_{\bf II}$\\
& $0.9373$ &$[110]_{\bf I}\to[010]_{\bf II}$&$[010]_{\bf I}||[102]_{\bf II}$\\
& $1.3424$ &$[001]_{\bf I}\to[001]_{\bf II}$&$[110]_{\bf I}||[010]_{\bf II}$\\
\end{tabular}
\end{ruledtabular}
 }
 \end{table}

\begin{acknowledgments}
We thank Liping Liu, Robert Kohn and Kaushik Bhattacharya for helpful discussions 
during the preparation of this work.
XC, YS, and RDJ acknowledge the support of the MURI project Managing
the Mosaic of Microstructure
(FA9550-12-1-0458, administered by AFOSR), NSF-PIRE (OISE-0967140) and 
ONR (N00014-14-1-0714). The Advanced Light Source is supported 
by the Director, Office of Science, Office of Basic Energy Sciences, 
of the U.S. Department of Energy under Contract No. DE-AC02-05CH11231.

\end{acknowledgments}

\bibliography{algorithm}

\end{document}